\documentclass[twocolumn,prl,superscriptaddress,showpacs,
amssymb,amsmath,amsfonts,nofootinbib]{revtex4}

\usepackage{graphicx} 
\usepackage{latexsym}
\usepackage{dcolumn} 
\usepackage{bm} 

\def\NC   {\rm Nuovo Cimento~}

\def\NIM  {\rm Nucl. Instr. and Meth.~}
\def\NIMA{{\rm Nucl. Instr. and Meth.} {\bf A}\,}

\def\PLB{{\rm Phys. Lett.} B\, }
\def\PR{{\rm Phys. Rev.~}}
\def\PRC{{\rm Phys. Rev.} C\,}
\def\PRD{{\rm Phys. Rev.} D\,}
\def\PRL{\rm Phys. Rev. Lett.~}

\newcommand{\etal}{{\em et al.}}

\begin{document}

\title{A New Measurement of the $\pi^0$ Radiative Decay Width}

\author{I.~Larin}
\affiliation{Alikhanov Institute for Theoretical and Experimental Physics, 
Moscow, Russia} 
\affiliation{North Carolina A\&T State University, Greensboro, NC 27411, USA}

\author{D.~McNulty}
\affiliation{Massachusetts Institute of Technology, Cambridge, MA 02139, USA}

\author{E.~Clinton}
\affiliation{University of Massachusetts, Amherst, MA 01003, USA}

\author{P.~Ambrozewicz}
\affiliation{North Carolina A\&T State University, Greensboro, NC 27411, USA}

\author{D.~Lawrence}
\affiliation{University of Massachusetts, Amherst, MA 01003, USA}
\affiliation{Thomas Jefferson National Accelerator Facility, 
Newport News, VA 23606, USA}

\author{I.~Nakagawa}
\affiliation{University of Kentucky, Lexington, KY 40506, USA}
\affiliation{RIKEN Nishina Center for Accelerator-Based Science, 
2-1 Hirosawa, Wako, Saitama 351-0198, Japan}

\author{Y.~Prok}
\affiliation{Massachusetts Institute of Technology, Cambridge, MA 02139, USA}

\author{A.~Teymurazyan}
\affiliation{University of Kentucky, Lexington, KY 40506, USA}

\author{A.~Ahmidouch}
\affiliation{North Carolina A\&T State University, Greensboro, NC 27411, USA}

\author{A.~Asratyan}
\affiliation{Alikhanov Institute for Theoretical and Experimental Physics, 
Moscow, Russia}

\author{K.~Baker}
\affiliation{Hampton university, Hampton, VA 23606, USA}

\author{L.~Benton}
\affiliation{North Carolina A\&T State University, Greensboro, NC 27411, USA}

\author{A.~M.~Bernstein}
\affiliation{Massachusetts Institute of Technology, Cambridge, MA 02139, USA}

\author{V.~Burkert}
\affiliation{Thomas Jefferson National Accelerator Facility, 
Newport News, VA 23606, USA}

\author{P.~Cole}
\affiliation{Idaho State University, Pocatello, ID 83209, USA}

\author{P.~Collins}
\affiliation{Arizona State University, Tempe, AZ 85287, USA}

\author{D.~Dale}
\affiliation{Idaho State University, Pocatello, ID 83209, USA}

\author{S.~Danagoulian}
\affiliation{North Carolina A\&T State University, Greensboro, NC 27411, USA}

\author{G.~Davidenko}
\affiliation{Alikhanov Institute for Theoretical and Experimental Physics, 
Moscow, Russia}

\author{R.~Demirchyan}
\affiliation{North Carolina A\&T State University, Greensboro, NC 27411, USA}

\author{A.~Deur}
\affiliation{Thomas Jefferson National Accelerator Facility, 
Newport News, VA 23606, USA}

\author{A.~Dolgolenko}
\affiliation{Alikhanov Institute for Theoretical and Experimental Physics, 
Moscow, Russia}

\author{G.~Dzyubenko}
\affiliation{Alikhanov Institute for Theoretical and Experimental Physics, 
Moscow, Russia}

\author{R.~Ent}
\affiliation{Thomas Jefferson National Accelerator Facility, 
Newport News, VA 23606, USA}

\author{A.~Evdokimov}
\affiliation{Alikhanov Institute for Theoretical and Experimental Physics, 
Moscow, Russia}

\author{J.~Feng}
\affiliation{University of North Carolina Wilmington, 
Wilmington, NC 28403, USA}
\affiliation{Chinese Institute of Atomic Energy, Beijing, China}

\author{M.~Gabrielyan}
\affiliation{University of Kentucky, Lexington, KY 40506, USA}

\author{L.~Gan}
\affiliation{University of North Carolina Wilmington, 
Wilmington, NC 28403, USA}

\author{A.~Gasparian}
\altaffiliation[Corresponding author:]{gasparan@jlab.org\\} 
\affiliation{North Carolina A\&T State University, Greensboro, NC 27411, USA}

\author{S.~Gevorkyan}
\affiliation{Yerevan Physics Institute, Yerevan, Armenia}
\affiliation{Joint Institute for Nuclear Research, Dubna, 141980, Russia}

\author{A.~Glamazdin}
\affiliation{Kharkov Institute of Physics and Technology, Kharkov, Ukraine}

\author{V.~Goryachev}
\affiliation{Alikhanov Institute for Theoretical and Experimental Physics, 
Moscow, Russia}

\author{V.~Gyurjyan}
\affiliation{Thomas Jefferson National Accelerator Facility, 
Newport News, VA 23606, USA}

\author{K.~Hardy}
\affiliation{North Carolina A\&T State University, Greensboro, NC 27411, USA}

\author{J.~He}
\affiliation{Institute of High Energy Physics, Chinese Academy of Sciences, 
Beijing, China}

\author{M.~Ito}
\affiliation{Thomas Jefferson National Accelerator Facility, 
Newport News, VA 23606, USA}

\author{L.~Jiang}
\affiliation{University of North Carolina Wilmington, 
Wilmington, NC 28403, USA}
\affiliation{Chinese Institute of Atomic Energy, Beijing, China}

\author{D.~Kashy}
\affiliation{Thomas Jefferson National Accelerator Facility, 
Newport News, VA 23606, USA}

\author{M.~Khandaker}
\affiliation{Norfolk State University, Norfolk, VA 23504, USA}

\author{P.~Kingsberry}
\affiliation{Massachusetts Institute of Technology, Cambridge, MA 02139, USA}
\affiliation{Norfolk State University, Norfolk, VA 23504, USA}

\author{A.~Kolarkar}
\affiliation{University of Kentucky, Lexington, KY 40506, USA}

\author{M.~Konchatnyi}
\affiliation{Kharkov Institute of Physics and Technology, Kharkov, Ukraine}

\author{A.~Korchin}
\affiliation{Kharkov Institute of Physics and Technology, Kharkov, Ukraine}

\author{W.~Korsch}
\affiliation{University of Kentucky, Lexington, KY 40506, USA}

\author{S.~Kowalski}
\affiliation{Massachusetts Institute of Technology, Cambridge, MA 02139, USA}

\author{M.~Kubantsev}
\affiliation{Alikhanov Institute for Theoretical and Experimental Physics, 
Moscow, Russia}
\affiliation{Northwestern University, Evanston/Chicago, IL 60208, USA}

\author{V.~Kubarovsky}
\affiliation{Thomas Jefferson National Accelerator Facility, 
Newport News, VA 23606, USA}

\author{X.~Li}
\affiliation{University of North Carolina Wilmington, 
Wilmington, NC 28403, USA}

\author{P.~Martel}
\affiliation{University of Massachusetts, Amherst, MA 01003, USA}

\author{V.~Matveev}
\affiliation{Alikhanov Institute for Theoretical and Experimental Physics, 
Moscow, Russia}

\author{B.~Mecking}
\affiliation{Thomas Jefferson National Accelerator Facility, 
Newport News, VA 23606, USA}

\author{B.~Milbrath}
\affiliation{Pacific Northwest National Laboratory, Richland, WA 99352, USA}

\author{R.~Minehart}
\affiliation{University of Virginia, Charlottesville, VA 22094, USA}

\author{R.~Miskimen}
\affiliation{University of Massachusetts, Amherst, MA 01003, USA}

\author{V.~Mochalov}
\affiliation{Institute for High Energy Physics, Protvino, Russia}

\author{S.~Mtingwa}
\affiliation{North Carolina A\&T State University, Greensboro, NC 27411, USA}

\author{S.~Overby}
\affiliation{North Carolina A\&T State University, Greensboro, NC 27411, USA}

\author{E.~Pasyuk}
\affiliation{Thomas Jefferson National Accelerator Facility, 
Newport News, VA 23606, USA}
\affiliation{Arizona State University, Tempe, AZ 85287, USA}

\author{M.~Payen}
\affiliation{North Carolina A\&T State University, Greensboro, NC 27411, USA}

\author{R.~Pedroni}
\affiliation{North Carolina A\&T State University, Greensboro, NC 27411, USA}

\author{B.~Ritchie}
\affiliation{Arizona State University, Tempe, AZ 85287, USA}

\author{T.~E.~Rodrigues}
\affiliation{University of S$\tilde{a}$o Paulo, S$\tilde{a}$o Paulo, Brazil}

\author{C.~Salgado}
\affiliation{Norfolk State University, Norfolk, VA 23504, USA}

\author{A.~Shahinyan}
\affiliation{Yerevan Physics Institute, Yerevan, Armenia}

\author{A.~Sitnikov}
\affiliation{Alikhanov Institute for Theoretical and Experimental Physics, 
Moscow, Russia}

\author{D.~Sober}
\affiliation{The Catholic University of America, Washington, DC 20064, USA}

\author{S.~Stepanyan}
\affiliation{Thomas Jefferson National Accelerator Facility, 
Newport News, VA 23606, USA}

\author{W.~Stephens}
\affiliation{University of Virginia, Charlottesville, VA 22094, USA}

\author{J.~Underwood}
\affiliation{North Carolina A\&T State University, Greensboro, NC 27411, USA}

\author{A.~Vasiliev}
\affiliation{Institute for High Energy Physics, Protvino, Russia}

\author{V.~Vishnyakov}
\affiliation{Alikhanov Institute for Theoretical and Experimental Physics, 
Moscow, Russia}

\author{M.~Wood}
\affiliation{University of Massachusetts, Amherst, MA 01003, USA}

\author{S.~Zhou}
\affiliation{Chinese Institute of Atomic Energy, Beijing, China}

\collaboration{PrimEx Collaboration}
\noaffiliation{}

\date{August 30, 2010}

\begin{abstract}
High precision measurements of the differential cross sections for $\pi^0$ 
photoproduction at forward angles for two nuclei, $^{12}$C and $^{208}$Pb, 
have been performed for incident photon energies of 4.9 - 5.5 GeV to extract 
the ${\pi^0 \to \gamma\gamma}$ decay width.
The experiment was done at Jefferson Lab using the Hall~B photon tagger and 
a high-resolution multichannel calorimeter. 
The ${\pi^0 \to \gamma\gamma}$ decay width was extracted by fitting the 
measured cross sections using recently updated theoretical models for the 
process. 
The resulting value for the decay width is 
$\Gamma{(\pi^0 \to \gamma\gamma)} = 7.82 \pm 0.14 ~({\rm stat.})
\pm 0.17 ~({\rm syst.}) ~{\rm eV}$. 
With the 2.8\% total uncertainty, this result is a factor of 2.5 more precise
than the current PDG average of this fundamental quantity and it is consistent 
with current theoretical predictions.

\end{abstract}


\maketitle

\begin{figure}[!ht]
\begin{center}
\includegraphics[scale=0.45]{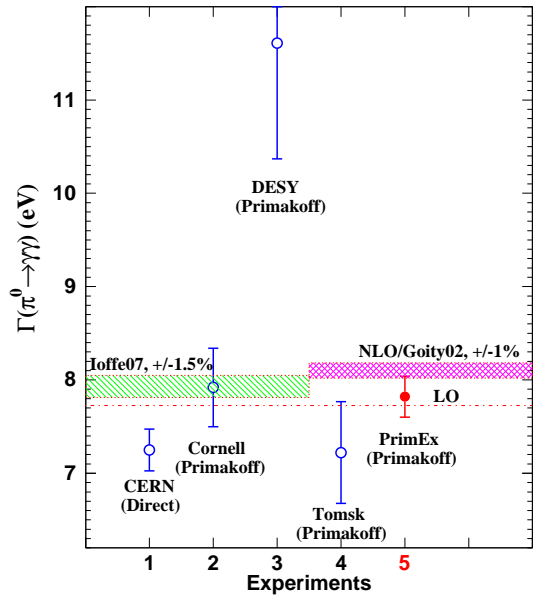}
\vspace{-0.50cm}
\caption{
$\pi^{0} \to \gamma \gamma$ decay width in eV. The dashed horizontal line is 
the LO chiral anomaly prediction.
NLO ChPT prediction~\cite{Goity02} is shown as the shaded band on r.h.s.
The l.h.s shaded band is the prediction from Ref.~\cite{Ioffe2007}.
The experimental results, included in the PDG average, are for: 
(1) done with the direct method~\cite{Ather85}, (2, 3, 4) with the Primakoff 
method~\cite{Cornell74,Bellettini70,Krysh70}, and (5) is the current 
PrimEx result.}
\label{PDG_data}
\end{center}
\end{figure}

The ${\pi^0 \to \gamma\gamma}$ decay represents one of the key processes in 
the anomaly sector of QCD. 
It provides the main test of the chiral anomaly~\cite{BellJackiw69,Adler69} 
and at the same time of the Nambu-Goldstone nature of the $\pi^0$ meson. 
The ${\pi^0 \to \gamma\gamma}$ decay amplitude is determined by the chiral 
anomaly resulting from the coupling of quarks to the electromagnetic field. 
In the limit of vanishing quark masses (chiral limit) the amplitude is 
exactly predicted and is expressed in terms of the fine structure constant, 
the $\pi^0$ decay constant, and the number of colors of 
QCD~\cite{BellJackiw69,Adler69}. 
In the real world there are corrections due to the non-vanishing quark masses. 
These corrections are primarily a result of state mixing effects in the 
$\pi^0$ meson, which result from the isospin symmetry breaking by 
$m_u<m_d$~\cite{Mous95,Goity02}.  
The corrections have been analyzed in the framework of Chiral Perturbation 
Theory (ChPT)~\cite{Mous95,Goity02,Mous2002,Mous2009} up to order $p^6$ 
(NLO in Fig.~\ref{PDG_data}), and shown to lead to an enhancement of about 
4.5\% in the $\pi^0$ decay width with respect to the case where state mixing 
is not included (LO in Fig.~\ref{PDG_data}). 
The estimated uncertainty in the ChPT prediction is 1\%~\cite{Goity02}.  
Corrections to the chiral anomaly have also been performed in the framework
of QCD using dispersion relations and sum rules~\cite{Ioffe2007}
(Ioffe07 in Fig.~\ref{PDG_data}). 
The fact that the corrections to the chiral anomaly are small and they are 
known at the 1\% level makes the ${\pi^0 \to \gamma\gamma}$ decay channel a 
benchmark process to test one of the fundamental predictions of QCD.

The current average experimental value for the $\pi^0$ decay width given by 
the Particle Data Group (PDG)~\cite{PDG2008} is 
$\Gamma{(\pi^0 \to \gamma\gamma)}=7.74 \pm 0.55 ~{\rm eV}$. 
This value is an average of four experiments with much larger dispersion 
between both the decay width values and their quoted experimental 
uncertainties, as shown in Fig.~\ref{PDG_data}. 
The most precise Primakoff type measurement was done at Cornell by 
Browman \etal~\cite{Cornell74} with a 5.3\% quoted total uncertainty: 
$\Gamma{(\pi^0 \to \gamma\gamma)}=7.92\pm0.42 ~ {\rm eV}$. 
This result agrees within experimental uncertainty with the theoretical 
predictions.  
Two other measurements~\cite{Bellettini70,Krysh70} with relatively large 
experimental uncertainties ($\simeq 11\%$ and $\simeq 7\%$) differ 
significantly from each other and do not agree with the theoretical 
predictions.
The most precise measurement of the $\pi^0$ decay width, prior to the current 
PrimEx experiment, was made by Atherton \etal~\cite{Ather85} using the direct 
method of measuring the mean decay length of $\pi^0$s produced by a high 
energy proton beam at CERN. 
Their result with the quoted 3.1\% total uncertainty, 
$\Gamma{(\pi^0 \to \gamma\gamma)}=7.25\pm0.18\pm0.14 ~ {\rm eV}$, is
$\sim 4 \sigma$ lower than the NLO ChPT prediction of Ref.~\cite{Goity02}. 

Clearly, a new Primakoff type experiment with a precision comparable to, or 
better than, the direct method measurement~\cite{Ather85} was needed to 
address the experimental situation on this fundamental quantity.

The PrimEx experiment~\cite{PrimEx} was performed in fall 2004 at the 
Thomas Jefferson National Accelerator Facility. 
It utilized the Hall~B high precision photon tagging 
facility~\cite{HallB-Tagger} together with a newly developed high resolution 
electromagnetic calorimeter.  
The combination of these two techniques greatly improved not only the angular 
resolutions, which are critical for Primakoff type measurements, but 
significantly reduced the systematic uncertainties that were present in 
previous experiments.
 
Tagged photons with known timing and energy were incident on two 5\% 
radiation length targets of $^{12}$C and $^{208}$Pb~\cite{target}. 
The photon tagging efficiencies were continuously measured during the 
experiment with a $e^{+}e^{-}$ pair spectrometer (PS) consisting of 
a $\sim$ 1.7~T$\cdot$m large aperture dipole magnet and two telescopes 
of scintillating counters located downstream of the targets. 
The absolute normalization of the photon beam was measured periodically 
with a total absorption counter (TAC) at low beam intensities.

The decay photons from $\pi^0\to\gamma\gamma$ were detected in a 
multichannel hybrid electromagnetic calorimeter (\mbox{HyCal}) located 7.5~m 
downstream from the targets to provide a large geometrical acceptance 
($\sim$70\%). 
\mbox{HyCal} consists of 1152 PbWO$_4$ crystal shower detectors 
($2.05 \times 2.05 \times 18.0$ cm$^3$) in the central part, surrounded by 
576 lead glass Cherenkov counters ($3.82 \times 3.82 \times 45.0$ cm$^3$). 
Four crystal detectors were removed from the central part of the calorimeter 
($4.1 \times 4.1$ cm$^2$ hole in size) for passage of the high intensity 
($\sim 10^7 ~\gamma/$s) incident photon beam through the 
calorimeter~\cite{HyCal-paper}. 
Twelve 5-mm-thick scintillator counters, located in front of HyCal, provided 
rejection of charged particles and effectively reduced the background in the 
experiment.
To minimize the decay photon conversion in air, the space between the PS 
magnet to HyCal was enclosed by a helium bag at atmospheric pressure. 
The photon beam's position stability was monitored during the experiment by 
an \mbox{X-Y} scintillating-fiber detector located downstream of 
\mbox{HyCal}. 

\begin{figure}[!hbt]
\vspace{-0.50cm}
\unitlength 1.0cm
\hskip -0.120truein
\begin{minipage}[th]{4.0cm}
\includegraphics[height=1.65\hsize, width=1.25\hsize]{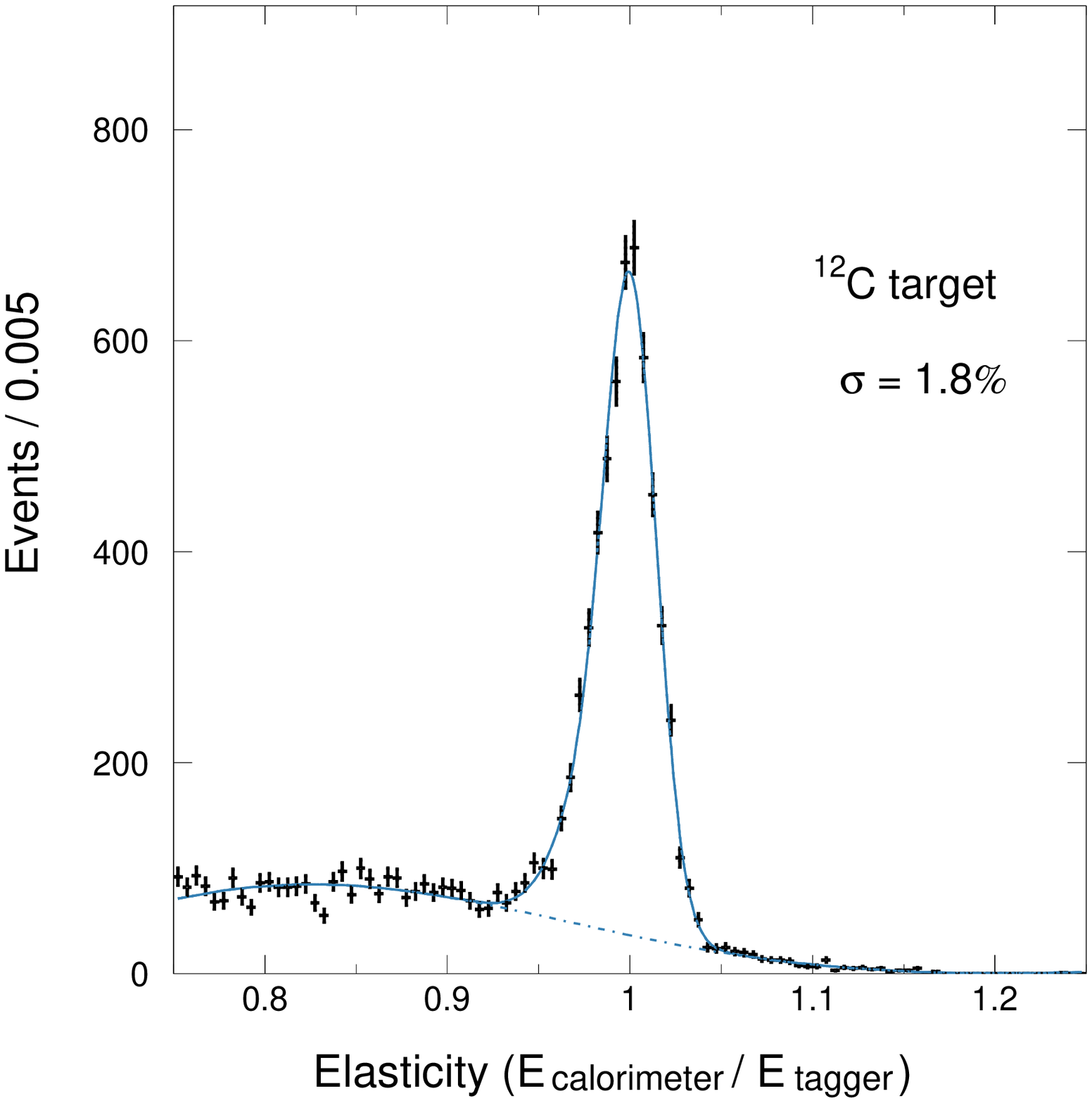}
\end{minipage}
\hskip 0.160truein
\begin{minipage}[th]{4.0cm}
\includegraphics[height=1.65\hsize, width=1.25\hsize]{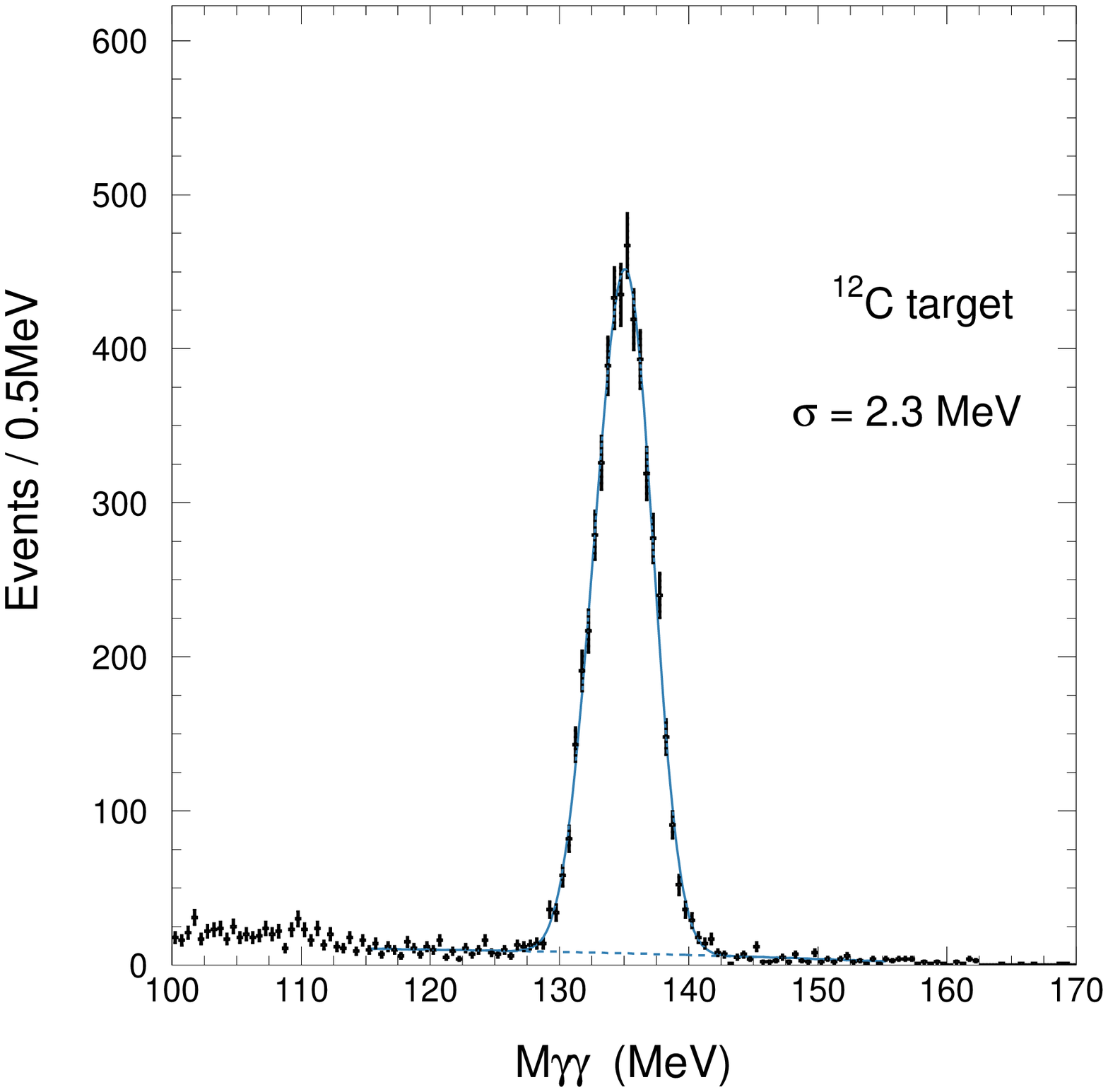}
\end{minipage}
\hfill
\vspace{-0.05cm}
\caption{Typical distribution of reconstructed ``elasticity'' (left panel) and 
$M_{\gamma\gamma}$ (right panel) for one angular bin.}
\label{data_elast_mgg}
\end{figure}

The experimental trigger was formed by requiring coincidences between the 
photon tagger in the upper energy interval (4.9 - 5.5 GeV) and HyCal with 
a total deposited energy greater than 2.5 GeV.
The combination of the photon tagger and the calorimeter defined the following 
main event selection criteria in this experiment: (1) timing between the 
incident photon and the decay photons in the calorimeter 
($\sigma_{t}$ = 1.1 ns); 
(2) ratio of the total energy in the calorimeter and the tagger energy, 
``elasticity'', ($\sigma_{el}$ = 1.8\%);
(3) invariant mass of the two photons ($M_{\gamma\gamma}$) reconstructed in 
the calorimeter (shown in Fig.~\ref{data_elast_mgg}). 

The event yield (number of $\pi^0$ events for each production angle bin) was 
obtained from the data by applying the selection criteria described above and 
fitting the experimental distributions of ``elasticity'' and $M_{\gamma\gamma}$
for each angular bin.
The typical background in the event selection process was only a few percent 
of the real signal events (see Fig.~\ref{data_elast_mgg}).
However, the uncertainty of 1.6\% in the background extraction in this much 
upgraded experiment still remained one of the largest contributions to the 
total systematic uncertainty.

The extraction of differential cross sections from the experimental yields 
requires an accurate knowledge of the total photon flux for each tagger energy 
bin, the number of atoms in the target, the acceptance of the experimental 
setup and the inefficiencies of the detectors. 
The uncertainty reached in the photon flux measurement, as described above, 
was at the level of 1\%~\cite{Photon-flux}. 
Different techniques have been used to determine the number of atoms in both 
targets with an uncertainty less than 0.1\%~\cite{target}. 
The acceptance and detection efficiencies and their uncertainties were 
calculated by a GEANT-based Monte Carlo code that included accurate 
information about the detector geometry and response of each detector 
element. 
Other than accidental backgrounds, some physics processes with an energetic 
$\pi^0$ in the final state can potentially contribute to the extracted 
yield.  
The $\omega$ photoproduction process through the $\omega \to \pi^0 \gamma$ 
decay channel is the dominant contribution to the background. 
The fit of the experimental data, as described below, with the subtracted 
physics background changes the extracted $\pi^0$ decay width by 1.4\% with 
an uncertainty of 0.25\%.

\begin{figure}[!ht]
\begin{center}
\includegraphics[scale=0.40]
{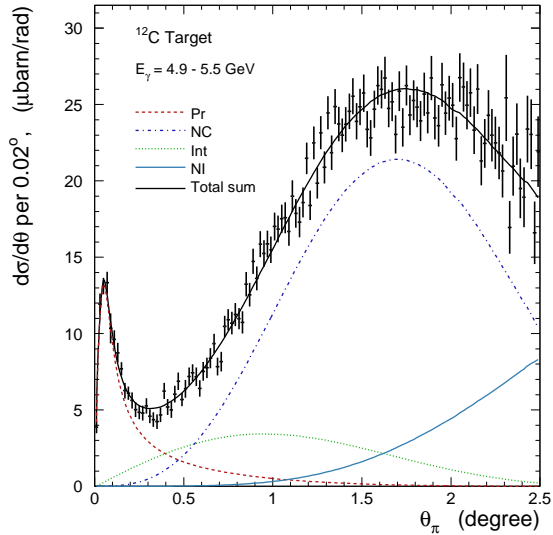}
\vspace{-0.35cm}
\caption{Differential cross section as a function of $\pi^0$ production angle 
for $^{12}$C together with fit ($\chi^2/N_{df}=152/121$) results for the 
different physics processes (see text for explanations).}
\label{crsect_12C}
\end{center}
\vspace{-0.65cm}
\end{figure}

\begin{figure}[!ht]
\begin{center}
\vspace{-0.50cm}
\includegraphics[scale=0.40]
{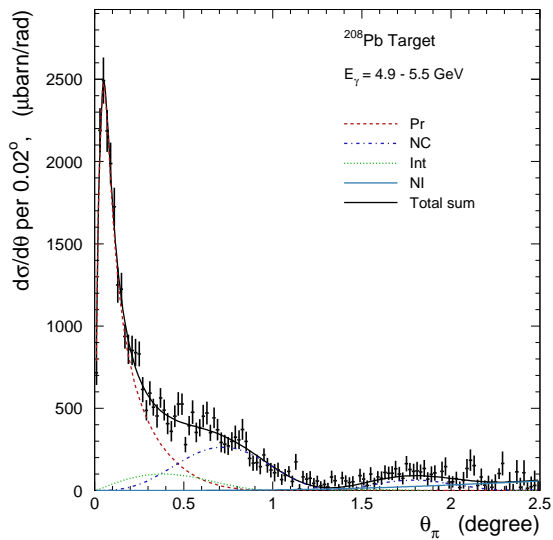}
\vspace{-0.35cm}
\caption{Differential cross section as a function of $\pi^0$ production angle 
for $^{208}$Pb together with fit ($\chi^2/N_{df}=123/121$) results for the 
different physics processes (see text for explanations).}
\label{crsect_Pb}
\end{center}
\vspace{-0.95cm}
\end{figure}

The resulting experimental cross sections for $^{12}$C and $^{208}$Pb are 
shown in Figs.~\ref{crsect_12C} and~\ref{crsect_Pb} along with the fit results 
for individual contributions from the different $\pi^0$ production mechanisms. 
Two elementary amplitudes, the Primakoff (one photon exchange), $T_{Pr}$, and 
the strong (hadron exchange), $T_S$, contribute coherently, as well as 
incoherently in $\pi^0$ photoproduction from nuclei at forward angles. 
The cross section of this process can be expressed by four terms: 
Primakoff ($Pr$), nuclear coherent ($NC$), interference between strong and 
Primakoff amplitudes  ($Int$), and nuclear incoherent ($NI$):
\vspace{-0.125cm}
\begin{eqnarray}
\frac{d\sigma}{d\Omega} &=& \mid{T_{Pr}+e^{i\varphi}T_S}\mid^2
+~\frac{d\sigma_{_{NI}}}{d\Omega} \nonumber \\
 &=& \frac{d\sigma_{_{Pr}}}{d\Omega}+\frac{d\sigma_{_{NC}}}{d\Omega}
~+\frac{d\sigma_{_{Int}}}{d\Omega}+\frac{d\sigma_{_{NI}}}{d\Omega}, \nonumber
\end{eqnarray}
where $\varphi$ is the relative phase between the Primakoff and the
strong amplitudes. 
The Primakoff cross section is proportional to the $\pi^0$ decay width, 
the primary focus of this experiment~\cite{Cornell74}:
\vspace{-0.125cm}
\begin{eqnarray}
\frac{d\sigma_{_{Pr}}}{d\Omega} = 
\Gamma{(\pi^0 \to \gamma\gamma)}\frac{8{\alpha}Z^2}{m^3}
\frac{\beta^3{E^4}}{Q^4}|F_{EM}(Q)|^2 \sin^2\theta_{\pi}, \nonumber
\end{eqnarray}
where $Z$ is the atomic number; $m$, $\beta$, $\theta_{\pi}$ are the mass, 
velocity and production angle of the pion; 
$E$ is the energy of the incident photon; 
$Q$ is the four-momentum transfer to the nucleus; 
$F_{EM}(Q)$ is the nuclear electromagnetic form factor, corrected for final 
state interactions (FSI) of the outgoing pion.
The FSI effects for the photoproduced pions, as well as the photon shadowing 
effect in nuclear matter, need to be accurately included in the cross sections 
before extracting the Primakoff amplitude.
To achieve this, and to calculate the $NC$ and $NI$ cross sections, a full 
theoretical description based on the Glauber method was developed, providing 
an accurate calculation of these processes in both light and heavy 
nuclei~\cite{Gevs-1,Gevs-2}. 
For the $NI$ process, an independent method based on the multi-collision 
intranuclear cascade model~\cite{Tulio-1} was also used to check 
the model dependence of the extracted decay width.
The uncertainty in the decay width from model dependence and the parameters 
inside of the models was estimated to be 0.3\%.
%
%

The $\Gamma{(\pi^0 \to \gamma\gamma)}$ decay width was extracted by fitting 
the experimental results with the theoretical cross sections of the four 
processes mentioned above folded with the angular resolutions 
($\sigma_{\theta_{\pi^0}}$ = 0.6 mrad) and the measured energy spectrum of 
the incident photons. 
In the fitting process, four parameters, 
$\Gamma{(\pi^0 \to \gamma\gamma)}$, $C_{NC}$, $C_{NI}$, $\varphi$, 
were varied to calculate the magnitude of the Primakoff, $NC$, $NI$ cross 
sections and the phase angle, respectively. 
Independent analyses of the experimental data by two groups within the PrimEx 
collaboration yielded the weighted averages of the extracted decay widths
for $^{12}$C and $^{208}$Pb presented in Table~\ref{tbl:fit_results}.
\begin{center}
\begin{table} 
\vspace{0.45cm}
\begin{tabular}{|c|c|c|c|c|c|}
\hline
\multicolumn{2}{|c|}{Target} & 
$\Gamma(\pi^0 \to \gamma\gamma)$ & $C_{NC}$ & $\varphi$ & $C_{NI}$ \\
\multicolumn{2}{|c|}{}       & 
[eV]                             &          & [rad]     &  \\
\hline
\multicolumn{2}{|c|}{$^{12}$C}   & 
  7.79$\pm$0.18 &  0.83$\pm$0.02 &   0.78$\pm$0.07 &   0.72$\pm$0.06 \\
\hline
\multicolumn{2}{|c|}{$^{208}$Pb} & 
  7.85$\pm$0.23 &  0.69$\pm$0.04 &   1.25$\pm$0.07 &   0.68$\pm$0.12 \\
\hline
\end{tabular}
\caption{The fit values extracted from the measured cross sections on 
$^{12}$C and $^{208}$Pb.
The values for the decay widths are the weighted averages from two analyses.
The uncertainties shown here are statistical only (see text for notations).
}
\label{tbl:fit_results}
\end{table}
\vspace{-0.95cm}
\end{center}
 
The extracted decay width combined for the two targets is 
$\Gamma{(\pi^0 \to \gamma\gamma)} = 7.82 \pm 0.14 ~({\rm stat.})
\pm 0.17 ~({\rm syst.})$~eV. 
The quoted total systematic uncertainty (2.1\%) is the quadratic sum of all 
the estimated uncertainties in this experiment. 
The systematic uncertainties were verified by measuring the cross sections 
of the Compton scattering and the $e^{+}e^{-}$ production processes. 
The extracted cross sections for these well-known processes agree with the 
theoretical predictions at the level of 1.5\% and will be published separately.
The PrimEx result, with a total experimental uncertainty of 2.8\%, is the most 
precise Primakoff type measurement of the $\Gamma{(\pi^0 \to \gamma\gamma)}$ 
to date. 
It is a factor of two-and-a-half more precise than the current average 
value quoted in the Particle Data Group for this important fundamental 
quantity. 
As a single experimental result, it directly confirms the validity of the 
chiral anomaly in QCD at the few percent level.
The goal of the PrimEx experiment has been to test the chiral anomaly
and the corrections to it in the $\pi^0$ decay width with high precision.
The second phase of this experiment is currently planned to run at 
Jefferson Lab to achieve the projected 1.4\% precision.

\vspace{0.15cm}
We acknowledge the invaluable contributions of the Accelerator and Physics 
Divisions at Jefferson Lab which made this experiment possible.
We thank the Hall~B engineering staff for their critical contributions in 
all stages of this experiment.
Theoretical support provided by Jose Goity throughout this project is 
gratefully acknowledged.

This project was supported in part by the National Science Foundation under 
a Major Research Instrumentation grant (PHY-0079840).
The Southern Universities Research Association (SURA) operated Jefferson Lab
under U.S. Department of Energy Contract No. DE-AC05-84ER40150 during this 
work.

\bibliographystyle{plain}

\end{document}